\documentclass[preprint]{revtex4}

\usepackage{graphicx}
\usepackage{dcolumn}
\usepackage{bm}
\usepackage{empheq}
\usepackage{color}
\usepackage{xcolor}
\usepackage{booktabs}
\usepackage{multirow}
\usepackage{amsmath}
\usepackage{tensor}

\begin{document}

\title{Vacuum fluctuation force on a rigid Casimir cavity in de Sitter and Schwarzschild-de Sitter spacetime}
\author{Xiang Chen}
 \affiliation{Arizona State University, Department of Physics, Tempe,
AZ 85287}
 \email{xchen48@asu.edu}   

\date{\today}

\begin{abstract}
We investigate the net force on a rigid Casimir cavity generated by vacuum fluctuations of electromagnetic field in three cases, de Sitter spacetime, de Sitter spacetime with weak gravitational field and Schwarzschild-de Sitter spacetime. In de Sitter spacetime the resulting net force follows the square inverse law but unfortunately it is too weak to be measurable due to the large universe radius. By introducing a weak gravitational field into the de Sitter spacetime, we find the net force now can be splited into two parts, one is the gravitational force due to the induced effective mass between the two plates, the other one is generated by the metric structure of de Sitter spacetime. In order to investigate the vacuum fluctuation force on the rigid cavity under strong gravitational field, we perform the similar analysis in Schwarzschild-de Sitter spacetime, results are obtained in three different limits. The most interesting one is when the cavity gets closer to the horizon of a blackhole, square inverse law is recovered and the repulsive force due to negative energy/mass of the cavity now has an observable strength. More important the force changes from being repulsive to attractive when the cavity crosses the event horizon, so that the energy/mass of the cavity switches the sign which suggests the unusual time direction inside the event horizon.
\end{abstract}

\maketitle

\section{Introduction}
Since Casimir discovered that there exists an attractive force between two parallel plates of infinite extent by considering the zero-point energy of the electromagnetic mode structure, ``Casimir effect" has become a very popular research field. In Casimir's original paper, the magnitude of such a kind of force in Minkowski spacetime is shown as \cite{Casimir}
\begin{equation}
F(r)=\frac{\pi^2\hbar c A}{240 d^4} \label{casimir}
\end{equation}
with $A$ the area of each plate and $d$ the separation. Without putting the parallel plates, the vacuum has a uniform energy density by assigning a zero-point energy of $\frac{1}{2}\hbar \omega$ to each electromagnetic mode which can be thought as a virtual photon since they are not excited and then are not on shell. But when two parallel plates are set up with a separation, the energy density of vacuum is not longer uniform and leads to a lower energy density within the two plates than that of outside  which results a net force of attraction. This effect makes people realize that zero-point fluctuation is real and has the directly physical sequence. Thorough studies have gone through both the theoretical and experimental side, detailed reviews are referred to\cite{SKLamoreaux,VV,FCapasso,KMilton,Lamoreaux2005}.\\
That Casimir energy gravitates is another part of recent theoretical discovery \cite{SAFulling,KAMilton}
and the vacuum fluctuation force on rigid Casimir cavity in weak gravitation field has also been studied in \cite{GBimonte2006,GBimonte2007,GBimonte2008,GBimonte2008-2,GBimonte2009} including spin-0 and spin-1 cases. An experiment has been suggested in \cite{ECalloni}to measure the Casimir force in Schwarzschild metric in order to show whether the virtual quanta from vacuum fluctuation satisfies principle of equivalence. \\
The above studies all embed weak gravitational field into Minkowski spacetime background, however as the astronomical evidence shows that we are actually living in an accelerated expansion universe, this gives a central place to the de Sitter geometry in cosmology\cite{NStraumann}.
On geometrical side, de Sitter spacetime and Minkowski spacetime both stem from the class of
Lorentzian manifolds. Being maximally symmetric, they admit kinematical symmetry
groups having ten generators\cite{AEinstein} and implies a constant curvature. In Minkowski spacetime, the constant is zero while in de Sitter spacetime it is a nonzero constant(either positive or negative depends on convention).
Although sharing the same features on geometrical side, they have a very different interpretation on physical side. Casimir effect in de Sitter space may be a clue to solve dark energy problem, a central and yet unsolved problem in fundamental physics and cosmology. \\
The effective lagrangian and energy-momentum tensor in de Sitter space is calculated for a scalar field in \cite{Dowker1976-1,Dowker1976} wherein a general zeta-function method is developed.
In \cite{Miao}, the Casimir energy of photon field in a static de Sitter space is calculated and it is proportional to the size of horizon with the same form as the holographic dark energy.
In \cite{MRSetare} the Casimir stress on two parallel plates in de Sitter space is found for massless scalar field by applying Robin boundary conditions on the plates. Result shows that false vacuum is formed between the two parallel plates and true vacuum formed outside, while the total Casimir force leads to an attraction of the plates, which is opposite to the result in \cite{Miao} for photon field implying a repulsive Casimir force. Besides, the force from the boundary term is in fourth power of inverse distance between the plates for massless scalar field. Fermionic Casimir effect has been studied in de Sitter spacetime \cite{Saharian2009,Saharian2011} as well as energy-momentum tensor and Casimir force for massive scalar field. \\
Motivated by these studies we evaluate the vacuum fluctuation force, i.e. the total casimir force on a Casimir cavity in de Sitter spacetime, de Sitter spacetime with a weak gravitational field and Schwarzschild-de Sitter spacetime, in order to find out how spacetime background affects the virtual quanta and then leads to the difference in the vacuum fluctuation force. Such a difference may open a way to verify which universe we are actually living in.

\section{The de Sitter Spacetime}\label{desitter}
In order to show the structure of the de Sitter spacetime, we start with a flat, five-dimensional spacetime $M_5$ with metric \cite{Miltutin}
\begin{eqnarray}
\label{metric}
ds^2=-(dx^0)^2+(dx^1)^2+(dx^2)^2+(dx^3)^2-(dx^5)^2
\end{eqnarray}
and set the convention
$\eta_{a,b}=(-,+,+,+,-) (a,b=0,1,2,3,5)$  and $ (\mu,\nu=0,1,2,3) $. Next
 embed a hypersphere $H_4$ with `radius' $a$ (the radius of universe) into this five-dimensional spacetime $M_5$ which is
\begin{eqnarray}
-\eta_{\mu\nu}x^{\mu}x^{\nu}+(x^5)^2=a^2  \label{H4}
\end{eqnarray}
This hypersphere $H_4$ is then called the de Sitter spacetime, a maximally symmetric subspace of $M_5$. From Eq.(\ref{H4}) we have
$ (dx^5)^2=(\eta_{\mu\nu}x^{\mu}dx^{\nu})^2/(x^5)^2$ and substitute it into Eq.(\ref{metric}). The line element on $H_4$ now becomes
\begin{eqnarray}
ds^2=\eta_{\mu\nu}dx^{\mu}dx^{\nu}+\frac{(\eta_{\mu\nu}x^{\mu}dx^{\nu})^2}{a^2+\eta_{mn}x^mx^n}
\end{eqnarray}
For any maximally symmetric space, the curvature is the same at each point. So that the metric, Christoffel connection and curvature tensor can be found by consider the vicinity of $x^{\mu}=0$ only,
\begin{eqnarray}
g_{\mu\nu}=\eta_{\mu\nu}+\frac{x_{\mu} x_{\nu}}{a^2}, \,\,\, \Gamma^{\mu}_{\nu\rho}=\frac{1}{a^2}x^{\mu} \eta_{\nu\rho}, \, \, \, R_{\mu\nu\rho\sigma}=\frac{1}{a^2}(\eta_{\mu\rho}\eta_{\nu\sigma}-\eta_{\mu\sigma}\eta_{\nu\rho})
\end{eqnarray}
and
\begin{equation}
g^{\mu\nu}=\eta^{\mu\nu}-\frac{x^{\mu}x^{\nu}}{a^2}\label{gtmunu}
\end{equation}
where $x^{\mu}=(-x_0,x_1,x_2,x_3) $ after dropping the higher order term.
The Ricci tensor is found to be proportional to the metric
\begin{equation}
R_{\mu\nu}=\frac{3}{a^2}g_{\mu\nu}\label{Ricci}
\end{equation}
leading to a constant scalar curvature
\begin{equation}
R=12/a^2  \label{scalarR}
\end{equation}
These are key ingredients in Einstein field equation with cosmological constant term \cite{Einstein1916},
\begin{equation}
R_{\mu\nu}+\Lambda g_{\mu\nu}-\frac{1}{2}R g_{\mu\nu}=\frac{8\pi G}{c^4} T_{\mu\nu} \label{EFE}
\end{equation}
where $\Lambda $  is the cosmological constant, $G$ is Newton's gravitational constant, $c$ is the speed of light in vacuum and $T_{\mu\nu}$ the stress-energy tensor or energy-momentum tensor. Substitute Eq.(\ref{Ricci})(\ref{scalarR}) into Eq.(\ref{EFE}), we can see immediately that de Sitter space is actually a vacuum solution to Einstein equation if the positive cosmological constant is set as
\begin{equation}
\Lambda =\frac{3}{a^2}
\end{equation}
The cosmological constant term can be treated as a part of energy-momentum tensor and put on the right side of Einstein equation, more detailed discussion can be found in\cite{Weinberg1989}.

\section{Vacuum fluctuation force in de Sitter space }
Suppose there are two identical parallel plates with proper area $A$ and proper separation $d$, one is placed at the origin $(0,0,0)$  and the other one is at $(0,0,d)$ with normal vector of the plates in $z$ direction. \\
In Minkowski spacetime, following the analysis of \cite{LSBrown}, the regularized energy-momentum tensor $\langle T^{\mu\nu}\rangle $ of quantum electrodynamics is
\begin{equation}
\langle T^{\mu\nu}\rangle =\frac{\pi^2 \hbar c}{180 d^4}(\frac{1}{4}g^{\mu\nu}- \hat{z}^{\mu}\hat{z}^{\nu})\label{emtensor}
\end{equation}
here $g^{\mu\nu}$ is the space-time metric and $\hat{z}^{\mu}=(0,0,0,1)$ is the unit space-like 4-vector in z-direction which is
orthogonal to the surface of two plates. This is the energy-momentum tensor of matter, without including the gravitational fields even if we generate this formula to curved space-time.
While in de Sitter spacetime(3+1 dimension), the enery-momentum tensor for a conformally coupled massless($\xi=1/6$) scalar field between two plates is given in\cite{Elizalde}
\begin{eqnarray}
\langle T^{\mu}_{\nu}\rangle &=& \langle T^{\mu}_{\nu}\rangle_{dS}+\frac{e^{-4t/a^2}}{6\pi^{2}}\text{diag}(1,1, 1, -3)
\times \int_0^{\infty} dx \frac{x^3}{c_1(x)c_2(x)e^{2dx}-1}  \label{desittensor}
\end{eqnarray}
where $ \langle T^{\mu}_{\nu}\rangle_{dS}$ is the renormalized vacuum expected value of energy-momentum tensor without plates. It was shown in \cite{Bunch1978} that $ \langle T^{\mu}_{\nu}\rangle_{dS}$ is proportional to the metric tensor of de Sitter spacetime which is
\begin{eqnarray}
\langle T^{\mu\nu}\rangle_{dS}= -\frac{1}{8\pi}(\frac{1}{6}-\xi) R g^{\mu\nu}
\end{eqnarray}
In Eq.(\ref{desittensor}), for Dirichlet boundary condition $c_i(x)=-1$ while for Neumann boundary condition $c_i(x)=1$ with $i=1,2$ represents the plate 1 and plate 2 respectively. For a conformally coupled massless scalar field with $\xi=1/6$, $\langle T^{\mu}_{\nu}\rangle_{dS}$ vanishes.
Notice that
\begin{eqnarray}
\text{diag}(1,1,1,-3)=4(\frac{1}{4}g^{\mu}_{\nu}- \hat{z}^{\mu}\hat{z}_{\nu})
\end{eqnarray}
Therefore, the energy-momentum tensor for conformally coupled massless scalar field can be rewritten as
\begin{eqnarray}
\langle T^{\mu\nu}\rangle &=& K(\frac{1}{4}g^{\mu\nu}- \hat{z}^{\mu}\hat{z}^{\nu})\label{simtensor}
\end{eqnarray}
where
\begin{equation}
 K=\frac{2e^{-4t/a^2}}{3\pi^{2}} \int_0^{\infty} dx \frac{x^3}{c_1(x)c_2(x)e^{2dx}-1} \nonumber
\end{equation}
Since according to \cite{Elizalde}, the electromagnetic field in $D=3$ is conformally invariant so that the Casimir problem
dealing with two perfectly conducting parallel plates can be reduced to the corresponding problem with two scalar modes with
Dirichlet boundary conditions and Neumann boundary conditions. Therefore the energy-momentum tensor for electromagnetic field
in de Sitter space would be twice of the one expressed in (\ref{simtensor}), which is
\begin{displaymath}
 K_{em}=2K =  \left\{
     \begin{array}{lr}
     \frac{\pi^2 \hbar c e^{-4t/a^2}}{180d^4} & : c_1(x)=c_2(x)=1,  -1 \\
     -\frac{7\pi^2 \hbar c e^{-4t/a^2}}{1440d^4} & : c_1(x)=-c_2(x)=1, -1
     \end{array}\label{Kem}
   \right.
\end{displaymath}
Here $c_1(x)=c_2(x)=1,-1$ represents that Dirichlet BCs or Neumann BCs are used for both plates, while for $c_1(x)=-c_2(x)=1,-1$, Dirichlet BCs is used for
one plate and Neumann BCs used for the other. Recall the age of universe $t\sim 10^{17} s$ and the radius $a\sim 10^{26}$, we can safely replace $e^{-4t/a^2}$ by 1 and omit it in the rest of our paper. The energy-momentum tensor of electromagnetic field can then be simplified as
\begin{eqnarray}
\langle T^{\mu \nu}\rangle_{em} &=& K_{em} (\frac{1}{4}g^{\mu\nu}- \hat{z}^{\mu}\hat{z}^{\nu})\label{desitemtensor}
\end{eqnarray}

It is worth to notice that the energy-momentum tensor of electromagnetic field here does not has vanishing divergence,
\begin{equation}
\nabla_{\mu}T^{\mu\nu}=\nabla_{\mu}g^{\mu\nu}+\nabla_{\mu} (\hat{z}^{\mu}\hat{z}^{\nu})=\nabla_{\mu} (\hat{z}^{\mu}\hat{z}^{\nu}) \ne  0
\end{equation}
This is not consistent with the usual covariant conservation of energy-momentum tensor for isolated system\cite{Dewitt1975}. A further consideration reveals that the Casimir cavity contains two parts, one is the vacuum fluctuation within the two plates including the energy-momentum tensor from Maxwell action, gauge-breaking term and ghost action(Eq.(2.8) in \cite{GBimonte2006}), the energy-momentum tensor of which is covariant conserved. The other part, i.e. $\hat{z}^{\mu}\hat{z}^{\nu}$ is the boundary set by the two plates, the divergence of which vanishes in Minkowski spacetime, but no longer in curved spacetime, and a net force will be induced on this system which is,
\begin{eqnarray}
\nabla_{\mu}T^{\mu\nu}=f^{\nu}.
\end{eqnarray}
However the two-plate cavity is an isolated system in curved spacetime, there should be no external force exerted on it besides the gravitation force which can be treated as a spacetime background. Therefore we have to modify this boundary term in order to have a vanishing divergence so that the covariant conservation of the energy-momentum holds, and the new boundary term should recover the boundary term $z^{\mu}z^{\nu}$ when going from general spacetime back to Minkowski spacetime. To meet the first requirement, we have to give up the terms like $\eta^{\mu\nu},n^{\mu}n^{\nu}, \eta^{\mu\rho}n_{\rho}n^{\nu}, n^{\mu}n_{\rho}\eta^{\rho\nu}$ with $n^{\mu}$ the unit direction vector, and the only terms left will be $g^{\mu\nu},g^{\mu 3}g^{\nu 3}$ due to the metric compatibility condition
\begin{eqnarray}
\nabla_{\mu}g^{\nu\rho}=0\label{compatibility}.
\end{eqnarray}
To meet the second requirement, we have to discard $g^{\mu\nu}$ since it reduces to $\eta^{\mu\nu}$ instead of $z^{\mu}z^{\nu}$ in Minkowski spacetime. So the only term that meets both requirement is $g^{\mu 3}g^{\nu 3}$ and it is easy to check that $g^{\mu 3}g^{\nu 3}\rightarrow \eta^{\mu 3}\eta^{\nu 3}=\hat{z}^{\mu}\hat{z}^{\nu}$ when going from general curved spacetime to Minkowski spacetime. Therefore we can rewrite the original energy-momentum in (\ref{desitemtensor}) as
\begin{eqnarray}
\langle T^{\mu \nu}\rangle_{em} &=& K_{em} (\frac{1}{4}g^{\mu\nu}- g^{\mu 3}g^{\nu 3})\label{newemtensor}.
\end{eqnarray}
Now it is easy to check that this new energy-momentum satisfies the covariant conservation, i.e. $\nabla_{\mu}T^{\mu\nu}=0$ according to the metric compatibility condition shown in Eq.(\ref{compatibility}). The correct form of this formula as the energy-momentum tensor in general curved space-time can be obtained from a more general consideration. The regularized and renormalized energy-momentum tensor of spin-1 field in curved spacetime was discussed in \cite{Christensen1978}, where the energy-momentum tensor for Maxwell field, gauge fixed field and ghost field were calculated via point-split method. For massless photons, the renormalized energy-momentum tensor can be expressed as (see Eq.(5.4)in \cite{Christensen1978})
\begin{eqnarray}
\langle T^{\mu \nu}\rangle_{\text{Maxwell}}+\langle T^{\mu \nu}\rangle_{\text{gauge}}+\langle T^{\mu \nu}\rangle_{\text{ghost}}&=&\lim_{x^{'}\rightarrow x}
\frac{1}{4}[\tensor{G}{_\lambda^{\lambda \mu\nu}}+\tensor{G}{_\lambda^{\lambda \nu\mu}}+\tensor{G}{^{\mu\nu\lambda}_\lambda}+\tensor{G}{^{\nu\mu\lambda}_\lambda}\nonumber\\
&&\hspace{7ex}-\tensor{G}{^{\lambda\mu\nu}_\lambda}-\tensor{G}{^{\lambda\nu\mu}_\lambda}-\tensor{G}{^{\mu\lambda}_\lambda^\nu}
-\tensor{G}{^{\nu\lambda}_\lambda^\mu}\nonumber\\
&&\hspace{7ex}-(\tensor{G}{_\lambda^\lambda_\rho^\rho}-\tensor{G}{_{\lambda\rho}^{\rho\lambda}})g^{\mu\nu}]\propto g^{\mu\nu}
\end{eqnarray}
which gives the same form as the first term in (\ref{newemtensor}) to the first order of metric. Here $G$ represents the Green's function. This also shows that the first term of (\ref{newemtensor}) arises from the contribution of Maxwell field, gauge fixed field and ghost field. While the second term arises from the boundary, and we find that only in that form it is able to satisfy the covariant conservation and reduce to the second term in (\ref{emtensor}) when going back from curved spacetime to Minkowski spacetime.

Following the formula given in (\ref{newemtensor}) and use the metric of de-sitter space in (\ref{gtmunu}), we find the explicit form of energy-momentum tensor in de Sitter space (all the calculations below are accurate to order $(x/a)^2$)
\begin{eqnarray}
\langle T^{\mu\nu}\rangle_0=\frac{K_{em}}{4}\left[\begin{array}{cccc}
-(1+\frac{x_0^2}{a^2}) & \frac{x_0x_1}{a^2} & \frac{x_0x_2}{a^2} &-\frac{3x_0x_3}{a^2} \\
\frac{x_0x_1}{a^2} & 1-\frac{x_1^2}{a^2}  & -\frac{x_1x_2}{a^2} & \frac{3x_1x_3}{a^2}  \\
\frac{x_0x_2}{a^2} &  -\frac{x_1x_2}{a^2} &1-\frac{x_2^2}{a^2}  &  \frac{3x_2x_3}{a^2} \\
\frac{-3x_0x_3}{a^2} & \frac{3x_1x_3}{a^2} & \frac{3x_2x_3}{4a^2}  &-3+\frac{7x_3^2}{a^2}
\end{array} \right] \label{Tmunu}
\end{eqnarray}
where $\langle \rangle_0 $ denotes the energy-momentum for vacuum.
It is easy to check the trace of energy-momentum tensor vanishes
\begin{equation}
T^{\mu}_{\mu}=g_{\mu\nu}T^{\mu\nu}=0
\end{equation}
the vanishing trace reflects the invariance of scale transformation, i.e. conformally invariant.
If $a\rightarrow \infty $, we recover the energy-momentum in Minkowski spacetime \cite{LSBrown}
\begin{equation}
T^{\mu\nu}=\frac{K_{em}}{4}(-1,1,1,-3)
\end{equation}

To find the vacuum fluctuation force density  we will however take the expression derived in \cite{CMoller}
\begin{eqnarray}
f_{\mu}=-\frac{1}{\sqrt{-g}}\frac{\partial}{\partial x^{\nu}}(\sqrt{-g}T^{\nu}_{\mu})+\frac{1}{2}\frac{\partial g_{\rho\sigma}}{\partial x^{\mu}} T^{\rho\sigma}\label{fmu}
\end{eqnarray}
where
\begin{equation}
g=det(g_{\mu\nu})=(x_0^2-x_1^2-x_2^2-x_3^2-a^2)/a^2
\end{equation}
 Substituting (\ref{Tmunu}) into (\ref{fmu}) we find the density of force
\begin{eqnarray}
f_{\mu} & =& f_{1\mu}+f_{2\mu}\nonumber\\
& = & \frac{K_{em}}{4a^2}(x_0,-x_1,-x_2,-9x_3)+\frac{K_{em}}{4a^2}(-x_0,x_1,x_2,-3 x_3)\nonumber\\
&=&(0,0,0,\frac{-3K_{em} x_3}{a^2})\label{fmu1}
\end{eqnarray}
where $f_{1\mu}, f_{2\mu}$ refer to the first and second term of Eq.(\ref{fmu}) respectively.
This shows that after the cancelation between $f_{1\mu}$ and $f_{2\mu}$, only the third component(which is z direction) has a non-vanishing value as we expect.
And the non-vanishing value is linear with $x_3 $ ($x_3=z$),
Integrate the force density Eq.(\ref{fmu1}) over the volumn, we find the $z$ component of the net force on rigid cavity,
\begin{equation}
F_z=A \int_0^d\sqrt{-g} f_z dz=\frac{-3 K_{em} d^2}{2a^2}= \left\{
     \begin{array}{lr}
     -\frac{\pi^2 A}{120 a^2}\frac{\hbar c}{d^2} & : c_1(x)=c_2(x)=1,  -1 \\
     \frac{7\pi^2 A}{960 a^2}\frac{\hbar c}{d^2} & : c_1(x)=-c_2(x)=1, -1
     \end{array} \label{integral}
   \right.
\end{equation}
here $A$ is the area of each plate. Obviously, the force is attractive when BCs is $c_1(x)=c_2(x)=1,-1$, i.e. Dirichlet or Neumann BCs on each plate, and is repulsive when BCs is $c_1(x)=-c_2(x)=1,-1$, i.e. Dirichlet on one plate and  Neumann BCs on the other. This is consistent with \cite{Elizalde} in scalar field case. Notice that the force here is proportional to $d^{-2}$, following the square inverse law instead of usually being inverse proportional to fourth power of the separation between two plates, like the expression in (\ref{casimir}). However due to its dependence of $a^{-2}$ and the very large radius $a$ of our universe, this force is too small to be measured and can actually be neglected, so that there is no significant net Casimir force on the rigid cavity. This result is consistent with the calculation in Eq.(67) of \cite{Elizalde}, where the two plates have the same magnitude of pressure but with opposite direction, leading to a zero net force on such a two-plate system.


\section{Vacuum Fluctuation Force with Weak Gravitation Field in de-Sitter Spacetime}
In the previous analysis, we consider the vacuum fluctuation from only de-sitter background without explicit gravitational field. If however the two parallel plates are located somewhere above the earth, the gravitational field at that point, though weak, may have some negligible effect on the quantum vacuum fluctuations of electromagnetic field in the region between the two plates. Assuming the gravitational field is in negative $z$-direction, the metric of space-time now becomes (similar way in\cite{CWMisner})
\begin{equation}
g_{\mu\nu}=\eta_{\mu\nu}+\frac{x_{\mu}x_{\nu}}{a^2}-2 A^{\rho}x_{\rho}\delta_{\mu 0}\delta_{\nu 0}\label{gmunu2}
\end{equation}
with $A^{\mu}=(0,0,0,g/c^2) $, $g$ is the magnitude of earth's gravitational acceleration. Since the gravitational field is weak, we only keep the first-order corrections to the line element and ignore the higher order\cite{ECalloni},
\[\langle T^{\mu\nu}\rangle_0=\frac{K_{em}}{4}\left[\begin{array}{cccc}
-(1+\frac{x_0^2}{a^2})+2A_3x_3 & \frac{x_0x_1}{a^2} & \frac{x_0x_2}{a^2} &-\frac{3x_0x_3}{a^2} \\
\frac{x_0x_1}{a^2} & 1-\frac{x_1^2}{a^2}  & -\frac{x_1x_2}{a^2} & \frac{3x_1x_3}{a^2}  \\
\frac{x_0x_2}{a^2} &  -\frac{x_1x_2}{a^2} &1-\frac{x_2^2}{a^2}  &  \frac{3x_2x_3}{a^2} \\
-\frac{3x_0x_3}{a^2} &  \frac{3x_1x_3}{a^2} &  \frac{3x_2x_3}{a^2}  &-3+7\frac{x_3^2}{a^2} \label{Tmunu2} \end{array} \right].\]
Following the same path as before, we find the density of force for weak gravitational field in de-sitter space, with the first term and second term in (\ref{fmu}) given as
\begin{eqnarray}
f_{1\mu} &=& \frac{K_{em}}{4a^2}\bigg(-x_0(1-3A_3x_3),\,\,-x_1(1+5A_3x_3),\,\,-x_2(1+5A_3x_3),\nonumber\\
& & \hspace{10ex}3A_3(x_0^2-x_1^2-x_2^2+a^2-3)-10A_3 x_3^2\bigg)\\
f_{2\mu}& = & \frac{K_{em}}{4a^2}\bigg(x_0(1+A_3x_3),\,\,x_1(1+A_3x_3),\,\,x_2(1+A_3x_3), A_3(x_0^2+x_3^2+a^2)-3x_3\bigg)
\end{eqnarray}
Add the two term together we have the net Casimir force density on the two plates system,
\begin{eqnarray}
f_{\mu} & =& f_{1\mu}+f_{2\mu}\nonumber\\
&=&\frac{K_{em}}{a^2}\bigg(x_0(A_3x_3),\,\,3x_1(A_3x_3)/2,\,\,3x_2(A_3x_3)/2,\nonumber\\
& &\hspace{10ex} A_3(a^2+x_0^2-3x_1^2/4-3x_2^2/4-9x_3^2/4)-3x_3\bigg)\nonumber\\
&\simeq &\bigg(0,\,\,0,\,\,0,\,\,K_{em} A_3-\frac{3K_{em} x_3}{a^2}\bigg)+\mathcal{O}((x/a)^2)\label{fmu2}
\end{eqnarray}
In the second to the last line we only keep the linear term of space-time coordinates and omit the higher order terms since they are small and negligible.
Integrate (\ref{fmu2}) over the volume as we show in(\ref{integral}) to find the net force that exerts on the two plates cavity,
\begin{equation}
F_z=K_{em} A(A_3d-\frac{3d^2}{2a^2})=F_{g}+F_{d}\label{netF1}
\end{equation}
where
\begin{equation}
F_g=K_{em}A A_3 d= \left\{
     \begin{array}{lr}
     \frac{\pi^2\hbar A }{180 c d^3}g & : c_1(x)=c_2(x)=1,  -1 \\
     -\frac{7\pi^2\hbar c A}{1440 c^2 d^3} g & : c_1(x)=-c_2(x)=1, -1
     \end{array}
   \right.
\end{equation}
\begin{equation}
F_d=-K_{em}A\frac{3d^2}{2a^2}= \left\{
     \begin{array}{lr}
     -\frac{\pi^2\hbar c A }{120 a^2 d^2} & : c_1(x)=c_2(x)=1,  -1 \\
     \frac{7\pi^2\hbar c A }{960 a^2 d^2} & : c_1(x)=-c_2(x)=1, -1
     \end{array}
   \right.
\end{equation}

Here $F_g$ as the first part of the Casimir force is proportional to the gravitational field strength $g$, therefore it can be considered as the gravitational force with effective mass $ \frac{-\pi^2\hbar A}{180 c d^3}$ which is in full agreement with Eq.(8) in \cite{ECalloni} and Eq.(5.4) in \cite{GBimonte2006} or $\frac{7\pi^2\hbar c A}{1440 c^2 d^3}$ depend on BCs. When both plates are in the same BCs, i.e. Dirichlet or Neumann BCs, the force $F_g$ orientates opposite to the gravitational field, so that the effective mass is considered to be negative. When two plates are in different BCs, the force $F_g$ follows the direction of gravitational field, so the effective mass should be positive.  The negative and positive effective mass reflects that the energy density between the two plates is smaller or larger than the outside respectively.

We also notice that the gravitational force is proportional to $d^{-3}$ and $\frac{1}{4}$ of $F_g$ is from the energy density $T^{00}$, while the rest is from the third component of energy-momentum tensor. According to the argument presented in \cite{ECalloni}, the net force that can be measured in experiment with BCs Dirichlet or Neumann for both plates is,
\begin{eqnarray}
F=\frac{1}{4}F_g=\frac{\pi^2\hbar A}{720 c d^3}
\end{eqnarray}
This reproduces Eq.(4.13) in \cite{GBimonte2007} and Eq.(9) in \cite{ECalloni} for a weak gravitational field in Minkowski space-time.

The second part of Casimir force in (\ref{netF1}) is $F_d$, which has no dependence on the gravitation field $g$, implying that this force is purely from the de-sitter spacetime. And $F_d$ is inverse proportional to the square of radius $a$ as well as the square of the separation $d$ between the two plates. Though the separation $d$ can be made very small in experiment, the radius of universe $a$ is so large that $a^2d^2$ is still large and can be ignored, we may expect $F_g$ to dominate the Casimir force $F_z$.
To confirm this, we calculate the critical value of $d$ which is $d_c=\frac{2 g a^2}{3 c^2}\approx 10^{35} m $ so that $F_g=F_d$ by taking $a \approx 10^{26}$ m and $ g\approx 10$ m$^2$/s. The critical value here is even larger than the present radius of our universe. Such an unreachable distance implies the first term dominates and the second term is negligible.
The Casimir force arises from de-Sitter space being negligible implies that the structure of our universe can not be determined by observing vacuum fluctuation of quantum electrodynamics in weak gravitational field. In order to see how things change in strong gravitational field, we will adopt the more general metric form i.e. Schwarzschild-de Sitter spacetime that can deal with strong gravitational field and perform similar analysis in the next section.

\section{Vacuum fluctuation force in Schwarzschild-de Sitter spacetime}
The static Schwarzschild-(anti-)de Sitter metric is expressed in spherical coordinates as\cite{Gibbons1977,Bousso1998,Podolsky1999,Cardoso2001,Cardoso2002} in signature (-, +,+,+),
\begin{equation}
ds^2=-(1-f)dt^2+\frac{1}{1-f}dr^2+r^2(d\theta^2+\sin^2\theta d\phi^2) \label{schwarz-desitter}
\end{equation}
with $f=(\frac{2GM}{c^2 r}+\frac{\Lambda r^2}{3}) $ where $M$ is the mass of the black hole or any spherical source of gravitational field. When
$\Lambda=0$, (\ref{schwarz-desitter}) reduces to Schwarzschild metric and when $M=0$ it reduces to de(anti-de) Sitter space.
As shown in \cite{Mackay} the metric tensor can be written in Cartesian coordinates $x=r\sin{\theta}\cos{\phi}, y=r\sin{\theta}\sin{\phi} $ and $z=r\cos{\theta}$ as
\begin{eqnarray}
g_{\mu\nu}=\left[\begin{array}{cccc}
-(1-f) & 0 & 0 & 0 \\
0 & 1+\frac{f x^2}{r^2(1-f)} & \frac{f x y}{r^2(1-f)} & \frac{f x z}{r^2(1-f)} \\
0 & \frac{f x y}{r^2(1-f)} & 1+\frac{f y^2}{r^2(1-f)} & \frac{f y z}{r^2(1-f)} \\
0 & \frac{f x z}{r^2(1-f)} & \frac{f y z}{r^2(1-f)} & 1+ \frac{f z^2}{r^2(1-f)}
\end{array}\right]
\label{s-dsitgmunu}
\end{eqnarray}
combine with Eq.(\ref{emtensor}) we have energy-momentum tensor
\begin{eqnarray}
T_{\mu\nu}=\frac{K_{em}}{12} \left[\begin{array}{cccc}
 -\frac{3}{1 - 2 \frac{G M}{c^2 r} - \Lambda r^2/3}, &  0 & 0 & 0 \nonumber\\
 0 & 3 -\frac{6 G M x_1^2}{c^2 r^3} - \Lambda x_1^2& - x_1 x_2(\frac{6 G M x_1^2}{c^2 r^3} + \Lambda) &
 - x_1 x_3 (\frac{6 G M x_1^2}{c^2 r^3} + \Lambda) \nonumber\\
 0 & - x_1 x_2 (\frac{6 G M x_1^2}{c^2 r^3} + \Lambda) & 3 -x_2^2 (\frac{6 G M x_1^2}{c^2 r^3} + \Lambda) &
 - x_2 x_3(\frac{6 G M x_1^2}{c^2 r^3} + \Lambda)  \nonumber\\
 0 & - x_1 x_3 (\frac{6 G M x_1^2}{c^2 r^3} + \Lambda) & - x_2 x_3 (\frac{6 G M x_1^2}{c^2 r^3} + \Lambda) &
  -9 - x_3^2 (\frac{6 G M x_1^2}{c^2 r^3} + \Lambda)
\end{array}\right]\\
\label{Sch-de-tensor}
\end{eqnarray}
It is worth to note here that the direct evaluation of the regularized and renormalized energy-momentum tensor in Schwarzschild-de sitter spacetime
with boundary remains a challenging task due to its complication in calculation, where the point-split method should be used for this purpose. Further evaluation will be considered in our future research.

Now substitute the energy-momentum tensor in Eq.(\ref{Sch-de-tensor}) into Eq.(\ref{fmu}) we finally obtain the Casimir force density
\begin{eqnarray}
f_{\mu} &=&
K_{em} \bigg\{-\frac{27 x_3}{r^4}\bigg[8 A_3^3 r^3 +
    4 A_3^2 r^2 (-2 + 3 \epsilon)+
    A_3r (3 - 8 \epsilon + 6 \epsilon^2)+ (-2 + \epsilon) \epsilon^2\bigg]\times \nonumber\\
&& \,\, \, \times (0,  x_1x_3,  x_2x_3,  r^2+x_3^2) +\frac{81 x_3}{r^2}(0,  0,  0, A_3 r-\epsilon )\bigg\}/(-3 + 6 A_3 r + 3 \epsilon)^3
\end{eqnarray}
where we define $A_3=\frac{GM}{c^2 r^2}, \epsilon=\frac{r^2}{a^2}$ and $a$ is related to $\Lambda $ through $\Lambda =3/a^2$.  The total force is obtained by integrate out the force density $f_{\mu}$, let the origin be in the center of the lower plate with length $L_1$ and width $L_2$, the two parallel plates are located at $(0,0,r)$ and $(0,0,r+d)$. We then have
\begin{eqnarray}
F &=& \int_r^{r+d}\int_{-L_2/2}^{L_2/2}\int_{-L_1/2}^{L_1/2} f_{\mu} dx_1 dx_2 dx_3 = (0, 0, 0, F_z)   \nonumber\\
F_z &=&\frac{K_{em} L_1 L_2}{(1- \delta)^5}\bigg\{ \frac{d}{r}\bigg[\frac{3 r^2}{a^2}+\frac{3}{2}\delta -7 \delta^2 + \frac{23}{2}\delta^3 -
8 \delta^4 +2 \delta^5 -\frac{r^2}{2a^2} (19\delta - 17 \delta^2 + 4 \delta^3)\bigg]  \nonumber\\
&&+\frac{d^2}{r^2}\bigg[\frac{3r^2}{2a^2}-\frac{3}{2}\delta + \frac{27}{4} \delta^2 -\frac{37}{4} \delta^3 + 5 \delta^4 -
\delta^5- \frac{r^2}{2a^2}(15 \delta - 24 \delta^2 + 6 \delta^3)\bigg]\bigg\}
\label{totforce}
\end{eqnarray}
here $ \delta = 2A_3r$ and far less than 1 when $r \gg r_c=\frac{2GM}{c^2} $ with $r_c$ is the horizon of a black hole, but it gets close to 1 when the Casimir cavity is near the horizon.

Now we consider several limits and for simplicity we only consider the BCs with both plates are in Dirichlet or Neumann. For the BCs with one plate in Dirichlet and the other in Neumann, the results are similar except for a factor difference including a negative sign.

(1)$ M \rightarrow 0 $ i.e. $ \delta \rightarrow 0 $, the Schwarzschild-de Sitter spacetime  then reduces to pure de Sitter spacetime, Eq.(\ref{totforce}) becomes
\begin{equation}
F_z = K_{em} A \frac{3d^2}{2a^2}= \frac{\pi^2 A}{120 a^2}\frac{\hbar c}{d^2}
\end{equation}
which is the same as Eq.(\ref{integral}) in pure de Sitter space under the BCs we have specified before.

(2) $ a \rightarrow \infty $, this reduces Schwarzschild-de Sitter spacetime to Schwarzschild spactime, and Eq.(\ref{totforce}) becomes
\begin{equation}
F_z =\frac{K_{em} A}{(1- \delta)^5}\bigg\{ \frac{d}{r}(\frac{3}{2}\delta -7 \delta^2 + \frac{23}{2}\delta^3 -
8 \delta^4 +2 \delta^5)+\frac{d^2}{r^2}(-\frac{3}{2}\delta + \frac{27}{4} \delta^2 -\frac{37}{4} \delta^3 + 5 \delta^4 -
\delta^5)\bigg\}
\end{equation}
For weak gravitation field limit, $\delta$ is small and only its first power should be kept so that
\begin{equation}
F_z =K_{em} A ( \frac{3d}{2r}\delta -\frac{3d^2}{2r^2})\delta \simeq K_{em} A \frac{3d}{2r}\delta = 3K_{em} A d A_3=\frac{\pi^2\hbar c A}{60 c^2 d^3}g
\label{netF2}
\end{equation}
which is three times larger as Eq.(\ref{netF1}) where we add uniform weak gravitation field into de Sitter space. At first sight, this seems to be inconsistent. However if we carefully compare the metric of spacetime in (\ref{gmunu2}) and (\ref{schwarz-desitter}), since in (\ref{gmunu2}) $g_{\mu\nu}=\eta_{\mu\nu}-2 A^{\rho}x_{\rho}\delta_{\mu 0}\delta_{\nu 0}$ as $a \rightarrow \infty$, while in (\ref{schwarz-desitter})
due to the extra contribution from $g_{rr}=1/(1-f) $ in Schwarzschild-de Sitter metric(isotropic but not uniform) which is different in metric (\ref{gmunu2}) where $g_{rr}=1 $, so that to the first order the metric in (\ref{schwarz-desitter}) can be written as\cite{GBimonte2008}
\begin{eqnarray}
ds^2=-(1+2A_{\rho}x_{\rho})dt^2+(1-2A_{\rho}x_{\rho})d\vec{r}^2
\end{eqnarray}
where the first term is exactly the same as that in (\ref{gmunu2}), but the second term contains an extra contribution which is $-2A_{\rho}x_{\rho}$, it is this extra term that leads to a factor of 3 in (\ref{netF2}). The essential difference lies in that we embed gravitational field in Minkowski spacetime in the last section, while here we are dealing with the Schwarzschild spacetime background when taking $a \rightarrow 0$.

(3). For strong gravitation field limit, we consider the two parallel plates located close to the horizon of a black hole that is $r\rightarrow r_c^{+}$ so that $\delta \rightarrow 1^{-}$, now we have
\begin{equation}
F_z =\lim_{\delta \rightarrow 1^{-}}\frac{K_{em} A}{(1- \delta)^5}\frac{3d^2}{a^2}=\lim_{\delta \rightarrow 1^{-}}\frac{1}{(1- \delta)^5}\frac{\pi^2\hbar c A}{60 a^2 d^2}
\end{equation}
As $(1-\delta)$ gets close to $0$, the net Casimir force on the two parallel plates system can be large enough to be observable and its direction is outwards which is similar to the case in\cite{GBimonte2008}, acting as a repulsive force between the plates and blackhole. The repulsive force is due to the fact that energy/mass of the cavity is negative. But following the equivalent principle, the cavity will experience a free falling and fall into the blackhole. The interesting part here is that the sign of the force will change when the cavity passes through the event horizon, i.e. it changes from repulsive to attractive, therefore the energy/mass of the cavity changes from negative to positive when crossing the event horizon. As energy and time are conjugate variables in quantum mechanics, the changes of the sign in energy may suggest the unusual direction of time flow inside the event horizon.

\section{Conclusions}\label{conclusions}
We have considered three cases of the vacuum fluctuation force on a rigid Casimir cavity with two parallel plates separated by a distance $d$ in de Sitter spacetime. In the first and simplest one, we only consider de Sitter space. By investigating on the energy-momentum tensor of electromagnetic field in de Sitter spacetime, we find out that there exists a net force on the rigid cavity which satisfies square-inverse law though too small to be measurable. Such a force is attractive when two plates are both in Dirichlet BCs or Neumann BCs, and it is repulsive when one plate is in Dirichlet BCs while the other one is in Neumann BCs.
In the second case, we introduce a weak uniform gravitational field in de Sitter spacetime, and find that the vacuum fluctuation force on the two parallel plates has two decoupled parts to the order of $x^2$, one($F_g$) is proportional to the weak gravitational field strength $g$ and has little dependence on the radius of universe, indicating that it is generated by the weak gravitational field acting on the effective mass induced by the difference in energy density between the region inside the two plates and the region outside the two plates. The effective mass is negative if we apply Dirichlet or Neumann BCs on both plates, and positive if one plate is applied Dirichlet BCs and the other one is applied Neumann BCs. The other part ($F_d$) of the force is from the vacuum fluctuation due to de Sitter spacetime structure only and is inverse proportional to the square of radius of universe, but independent on the weak gravitational field. Due to the large radius of the present universe, the first part $F_g$ dominates this vacuum fluctuation force and it is measurable with a large area of each plate and a small gap between them even though the gravitational field strength as small as that on the earth.


Finally we study the vacuum fluctuation force in Schwarzschild-de Sitter spacetime, a more general spacetime structure so that strong gravitation field can be included. After the final result is obtained we take several limits, in zero gravitational field limit we recover the result in pure de Sitter space; in Schwarzschild spacetime limit with weak gravitation field we can reproduce the Casimir force in Minkowski spacetime with gravitation field except for a factor difference; in strong gravitational field limit the net Casimir force on rigid cavity near event horizon is found to be inverse proportional to the square of the separation between two parallel plates and the square of radius of universe, again follows the square inverse law. The force is repulsive and becomes larger as the cavity gets closer to the event horizon of blackhole. When the cavity crosses the event horizon, the repulsive force switches into attractive force, leading to the change from negative energy to positive energy within the cavity. This implies the unusual direction of time flowing inside the event horizon.


\end{document}